\documentclass[
pre,
twocolumn,
twoside,
byrevtex,
superscriptaddress,
floatfix
]{revtex4-2}

\usepackage{currfile}
\usepackage[justification=raggedright,singlelinecheck=false]{caption}

\usepackage{xurl}

\PassOptionsToPackage{hyphens}{url}\usepackage{hyperref}
\makeatletter
\g@addto@macro{\UrlBreaks}{\UrlOrds}
\makeatother

\usepackage{colortbl}

\definecolor{goodblue}{RGB}{0, 91, 187}

\usepackage{hyperref}
\hypersetup{
  hypertexnames=false,
  colorlinks=true,
  allcolors=goodblue,
  urlcolor=goodblue,
  linkcolor=goodblue,
  citecolor=goodblue,
  pdfborder={0 0 0},
  breaklinks=true,
}

\usepackage{graphicx,epsfig,verbatim,enumerate}
\usepackage{amssymb,amsmath}
\usepackage{ifthen}
\usepackage{float}

\usepackage{placeins}

\usepackage{mathtools}
\usepackage{booktabs}
\newboolean{twocolswitch}

\usepackage{subcaption}

\newcommand{\sindex}[1]{}
\newcommand{\nindex}[1]{}

\newcommand{\www}[1]{\url{#1}}

\usepackage{lettrine}

\usepackage{color}

\setboolean{twocolswitch}{true}

\begin{document}

%% note: article titles are lowercased except for proper nouns etc.

\title{\protect
The triumphs and tragedies of fandom: 
Emotional arcs in NFL tweets
}

\author{Elisabeth Kollrack}
\email{elisabeth.kollrack@uvm.edu}
\affiliation{Department of Mathematics \& Statistics, University of Vermont, Burlington, VT 05405}
\affiliation{Computational Story Lab, University of Vermont, Burlington, VT 05405}

\author{Michael V. Arnold}
\affiliation{Computational Story Lab, University of Vermont, Burlington, VT 05405}
\affiliation{Vermont Complex Systems Institute, University of Vermont, Burlington, VT 05405}

\author{Peter Sheridan Dodds}

\affiliation{Computational Story Lab, University of Vermont, Burlington, VT 05405}
\affiliation{Vermont Complex Systems Institute, University of Vermont, Burlington, VT 05405}
\affiliation{Vermont Advanced Computing Center, University of Vermont, Burlington, VT 05405}
\affiliation{Department of Computer Science, University of Vermont, Burlington, VT 05405}
\affiliation{
  Santa Fe Institute,
  1399 Hyde Park Rd,
  Santa Fe,
  NM 87501,
  US
}

\affiliation{
  Complexity Science Hub,
  Metternichgasse 8,
  1030 Vienna,
  Austria
}

\author{Christopher M. Danforth}
\email{chris.danforth@uvm.edu}
\affiliation{Department of Mathematics \& Statistics, University of Vermont, Burlington, VT 05405}
\affiliation{Computational Story Lab, University of Vermont, Burlington, VT 05405}
\affiliation{Vermont Complex Systems Institute, University of Vermont, Burlington, VT 05405}
\affiliation{Vermont Advanced Computing Center, University of Vermont, Burlington, VT 05405}

\date{\today}

\begin{abstract}
  \protect
\setlength{\parindent}{0pt}
\setlength{\parskip}{0.6\baselineskip plus 0.2\baselineskip minus 0.1\baselineskip}

Online fandom communities influence public opinion toward movies, musicians, and sports teams. 
Using a corpus of game-referencing tweets,
we measure variation in sentiment toward National Football League (NFL) teams driven by geography, game outcomes, and team performance for the 2011--2014 NFL seasons.
We estimate a fandom radius for each team, identifying regions where engagement exceeds background levels of discussion. 
We find sentiment for both winning and losing teams is positive
immediately prior to games,
drops at kickoff, 
and rebounds slightly during halftime. 
After halftime however, the trajectories diverge: 
Sentiment for winning teams increases toward the end of the game, while sentiment for losing teams remains low,
though both end up below their start of game levels.
Finally, a comparison between sentiment and win percentage 
reveals a weak positive relationship, suggesting that while team success contributes to fandom happiness, other factors also influence how fans discuss the NFL on social media. 
Our work contributes to a growing body of computational social science research that quantifies the many aspects
of modern fandom.

\end{abstract}

%% deprecated by the APS a while ago!
%% \pacs{89.65.-s,89.75.Da,89.75.Fb,89.75.-k}

\begingroup
\setlength{\parskip}{0pt}%
\maketitle
\endgroup

\section{Introduction}

%% \done{All figures should have a single includegraphics; combine figures into single pieces where needed; add labels (A, B, C) directly in figures if not already there; captions all go into one caption at the bottom.}

%% \done{Figures in the appendix have to be directly referenced in the main text. 
%% These ones are missing:}

Mobile devices have allowed fans to engage with professional sports without being physically present at games \cite{Khanna2025}. Through digital platforms and social media, fans can interact and react with one another in real time, sharing moments of joy, frustration, and record-breaking achievements \cite{Young2017}. For example, Twitter provides time-stamped and geo-tagged text, making it a valuable source for studying fan engagement \cite{Green2024}. More recently, the proliferation of sports gambling and prediction markets has further driven online engagement. Twitter allows sports bettors to exchange information, react to live events, and influence one another's wagering behavior \cite{Purdum2022}.

Prior work has used Twitter data to describe and explain the dynamics of collective attention across vastly different types of events. Arnold et al. \cite{Arnold2021Hurricanes} examined tweets involving the names of hurricanes and compared public attention on social media to hurricane severity. The key findings were that stronger hurricanes generated more engagement regardless of death toll, and that storms affecting the continental United States received substantially more attention than those impacting regions such as Puerto Rico. Their results suggest that Twitter engagement can influence disaster response and the allocation of funding and aid.

Twitter is also used for real-time event detection. Sakaki et al. \cite{Sakaki2010} created a Support Vector Machine to track tweets based on keywords and context as an early indicator of events. Filters were also used to track the location and severity of events. Event tracking could be helpful in the present research to determine which team the author of a tweet likely supports. When implemented in Japan, earthquakes of seismic intensity 3 or higher were detected by the algorithm. In some cases, this system detected earthquakes faster than the Japan Meteorological Agency, taking advantage of social media users as a proxy for seismometer sensors.

Twitter is a key platform for social media engagement in the context of sports. Einsle et al. examined how user interactions such as likes, comments, and shares boost fan engagement \cite{Einsle2023}. By enabling retweets and replies, Twitter allows fans to interact with one another in real time and quickly share updates about games. Social media allows fans to remain connected and avoid missing major events if they are not watching a game live. Fear of Missing Out (FoMO) leads to Continuous Partial Attention (CPA), where fans interact on social media while watching games \cite{LeeNa2024}. Twitter is especially useful for aggregating discussions about sporting events because hashtags help organize conversations and increase the visibility of trending topics.

Smith researched differences between fans who tweet during games and those who tweet after games \cite{Smith2019}. The study explores how Twitter use affects fans’ enjoyment of events and the role of Twitter as a second screen. The authors found that fans who used Twitter while watching games appeared to enjoy the experience more than those who only read tweets. They also found that fans who interacted with others online were more interested in the game and reported higher enjoyment. These results suggest that Twitter interaction provides fans with a sense of community, which enhances enjoyment.

Geographic identity also plays a role in how fans maintain loyalty to teams outside their local market. Reifurth et al. surveyed 920 nonlocal U.S. sports fans and found that physical distance from a team did not weaken team identification \cite{Reifurth2019}. Fans who identified with the city of their favorite team showed stronger loyalty than those who identified with a broader region. These results suggest that fandom is shaped not just by proximity, but by psychological connections to place. For this research, geotagged tweet distributions may reflect these geographic identity patterns, helping explain why certain teams attract fan activity far from their home markets.

A study by Chierichetti et al. \cite{Chierichetti2014} 
used real-time event data to examine how large populations responded to major events such as the Super Bowl \cite{Linnell2021}, award shows, and the World Cup. These events were detected using tweet volume and recent activity. The authors found that during games, original tweets increase while replies and retweets decrease. After games, the opposite pattern emerges, with replies and retweets becoming more common. The researchers used a logistic regression classifier to detect events and a probabilistic model to differentiate between general tweeting and event based tweeting. For our research, it is important to ensure that tweets referring to the game are tracked rather than all tweets posted during the game.

Corney \cite{Corney2014} conducted similar research and used word frequency to identify popular topics. The study focused on FA Cup finals and examined significant moments during matches. Their system used clustering to detect important moments, including those occurring during overlapping events. The authors were able to distinguish between different teams’ fans and how they reacted differently to the same event. Their methods successfully detected most major events, such as goals and saves, during World Cup finals.

Twitter data has also been used to measure societal happiness over time. Dodds et al. \cite{Dodds2011} created the Hedonometer, a tool that assigns happiness scores based on the words used in tweets. The Hedonometer found that Saturday is the “happiest” day of the week and Tuesday the “saddest.” Holidays tend to be associated with higher happiness scores, while natural disasters and celebrity deaths correspond to decreases. Words such as ``love” and ``food” are associated with higher happiness, while words such as ``hate” and ``funeral” are associated with lower happiness. The Hedonometer can be used as a real-time tracker of population level mood, as well as for event detection and analysis of public responses. This instrument is especially helpful for examining how fandom reactions differ based on team success.

Reagan et al. used Hedonometer to analyze emotional arcs in a corpus of fiction from Project Gutenberg \cite{Reagan2016}. To construct these arcs, the authors divided each story into 10,000 word segments and assigned a happiness score to each segment based on the words it contained. These scores formed a time series representing the emotional trajectory of the story. They applied three independent methods for this analysis: Singular Value Decomposition (SVD), hierarchical clustering, and a neural network. Notably, all three approaches converged on the same set of results, strengthening their findings. The analysis revealed six fundamental emotional arc patterns that characterize the majority of stories. One of these, the ``Man in a Hole” arc, consists of a decline in sentiment followed by a recovery, and is particularly relevant to our research, as discussed later in this paper.

Storywrangler is a natural language processing tool developed to track sociolinguistic and cultural trends \cite{Alshaabi2021Storywrangler}. The system analyzes over 100 billion tweets and allows one to explore word popularity, cultural shifts, and social amplification. Tweets are broken up into n-grams (sequences of words) and analyzed by frequency, rank, and popularity as measured by retweets. Our research uses a similar approach, as it searches for Twitter hashtags (team names) and tracks their frequencies through time. Storywrangler has been used previously to analyze social movements (\#BlackLivesMatter, \#MeToo), politics, COVID-19, and popular culture, providing useful insights into human communication dynamics\cite{Wu2023, Weaving2023, Fudolig2022, Alshaabi2021Covid, Stupinski2022}.

Fandom often extends beyond a team’s physical location. The present study uses social media to estimate the geographic extent of sports fandom empirically, which we refer to as the ``fandom radii''. Rather than relying on traditional measures such as ticket sales or surveys, we use geotagged tweets to identify where fans are located. This study analyzes how sentiment differs across fandoms as well as between winning and losing teams during games.

Together, these studies demonstrate that Twitter data can be used to detect events, measure attention, and infer collective behavior in real time. Building on this literature, the following section outlines the data and methods used to identify and analyze geographic patterns of sports fandom using geotagged tweets.

\begin{table*}[tp!]
\caption{Example calculations of fandom radius components for selected NFL teams. For each team, the table reports normalized tweet activity at the city center $A(M_1)$, the index $k$ at which the 70\% cumulative population threshold is reached, the corresponding baseline distance $d_{\text{baseline}}$, the number of metropolitan areas $N$ used to compute baseline activity, the baseline activity level $A_{\text{baseline}}$, and the resulting fandom radius $R_{\text{fandom}}$.}
\label{tab:radii}
\begin{ruledtabular}
\begin{tabular}{lcccccc}
Team & $A(M_1)$ & $k$ & $d_{\text{baseline}}$ (km) & $N$ & $A_{\text{baseline}}$ & $R_{\text{fandom}}$ (km) \\
\hline
Seattle Seahawks & 56 & 250 & 3738 & 55 & 0.92 & 782 \\
Dallas Cowboys & 29 & 266 & 1975 & 86 & 3.03 & 369 \\
New England Patriots & 29 & 241 & 2578 & 83 & 1.6 & 232 \\
Baltimore Ravens & 45 & 213 & 2011 & 73 & 0.94 & 121 \\
\end{tabular}
\end{ruledtabular}
\end{table*}

\section{Methods}

Twitter data used for this project were collected through the Vermont Advanced Computing Center (VACC). Twitter’s Decahose API provides access to a random 10\% sample of all public tweets from September 2008 through June 2023. Each tweet includes metadata such as the tweet text, timestamp, and geographic coordinates when available.

The dataset for this study consists of tweets authored during the 2011 through 2014 National Football League (NFL) regular seasons. The 2011 season began on September 8, 2011, and the 2014 regular season concluded on December 28, 2014. Only tweets authored during the regular season are included in the dataset. We chose this time period based on the fact that geo-tagged messages, a functionality that became available in 2010, became less common following design changes in the Twitter mobile app in 2015.

Tweets were collected surrounding each NFL game played during the study period (2011-2014). For every game involving a given team, a seven day window is constructed spanning three days before the game through three days after the game. Tweets were retrieved if they occurred within this window and contained team related hashtags.
Tweet timestamps were recorded in Coordinated Universal Time (UTC), while official game kickoff times were listed in Eastern Time (ET). To ensure consistency, all game times were converted to UTC using Python’s \texttt{pytz} package prior to constructing the time windows.

Hashtags were defined for each team using a predefined set of team identifiers, including team mascots (e.g., \#Bills) and full team names (e.g., \#BuffaloBills; see Appendix Table \ref{tab:hashtags}). In addition, matchup specific hashtags were generated for each game using both team abbreviations (e.g., \#NEvsBUF and \#BUFvsNE). Hashtag matching is performed in a case insensitive manner.

Tweets were retrieved separately for each hashtag and then combined into a single dataset. Duplicate tweets resulting from multiple hashtag matches were removed using unique tweet identifiers. For each tweet, metadata were recorded including the associated team, their opponent, game identifier, season, home/away designation, and game timestamp.

Metropolitan population data were obtained at the Core Based Statistical Area (CBSA) level using annual population estimates from 2011 through 2014. The dataset is restricted to Metropolitan Statistical Areas. Season specific population estimates were used for analyses conducted at the yearly level.
For analyses aggregating across all seasons, the average population across the four years is calculated and used as the population measure. Population estimates were merged with a CBSA shapefile using CBSA identifiers, allowing population data to be spatially joined with the metropolitan regions used in the geographic analysis.

To quantify the geographic extent of each NFL team’s fanbase, we define and compute a \emph{fandom radius} based on the spatial distribution of geolocated Twitter activity. Each geo-tagged tweet is associated with latitude and longitude coordinates, and distances are measured relative to a team specific city center.
For consistency across teams, the city center is defined using the primary city, often referenced in the team’s name, rather than the physical location of the stadium. For example, the New England Patriots play their home games in Foxborough, Massachusetts, but the geographic center of Boston is used as the team’s reference point. This choice reflects the broader metropolitan identity of each franchise rather than stadium specific geography.

Tweets are spatially joined to U.S. Core Based Statistical Areas (CBSAs), representing major metropolitan regions, using \texttt{geopandas}. Distances between tweet locations and team city centers are computed using a projected coordinate system (EPSG:5070) to ensure accurate measurements in kilometers. For each metropolitan area, we calculate the mean distance of its associated tweets from the city center as well as the total number of tweets originating within that region.
Metropolitan areas vary substantially in population size, and so raw tweet counts are normalized by population to allow accurate comparison across regions. Tweet activity is expressed as tweets per 100,000 residents using season specific metropolitan population estimates. 
Normalized tweet activity for each metropolitan area, $A(M_i)$, is defined as follows:

\begin{align}
A(M_i) &= \frac{\text{Tweets}(M_i)}{\text{Population}(M_i)} \times \textnormal{100,000}
\label{eq:metro_area}
\end{align}

Metropolitan areas are then ordered by increasing distance from the team’s city center, and their populations are cumulatively summed. The metropolitan areas are represented by \(M_1, M_2, \dots, M_n\) and are sorted by increasing distance \(d_i\) from the team's city center. We define a core region as the set of metropolitan areas whose combined population accounts for at least 70\% of the total population represented in the data. The baseline distance, $d_{\text{baseline}}$, is defined as the distance of the metropolitan area at which this 70\% population threshold is first reached. Formally, this distance is given by:

\begin{align}
\begin{split}
d_{\text{baseline}} &= d_k, \\
k &= \min \left\{ m : \sum_{i=1}^m P_i \ge 0.70
\sum_{i=1}^n P_i \right\}
\end{split}
\label{eq:baseline_distance}
\end{align}

 Here, $k$ represents the index of the metropolitan area at which the cumulative population first reaches at least 70\% of the total population. In other words, $M_k$ is the outermost metropolitan area included in the core region when aggregating outward from the team’s city center. $d_{\text{baseline}}$ is intended to capture a team’s primary regional market.

Baseline tweet activity, $A_{\text{baseline}}$, is calculated as the average normalized tweet activity across all metropolitan areas located at distances greater than or equal to the baseline distance. Baseline activity is defined as: 
\begin{align}
A_{\text{baseline}} &= \frac{1}{N} \sum_{i: d_i \geq d_{\text{baseline}}} A(M_i)
\label{eq:baseline_act}
\end{align}

 where $N = |\{i : d_i \geq d_{\text{baseline}}\}|$ is the number of metropolitan areas at or beyond the baseline distance. Metropolitan areas with $d_i \geq d_{\text{baseline}}$ are treated as the baseline (background) region used to estimate typical non-local engagement. The \textit{fandom radius}, $R_{\text{fandom}}$, is then defined as the distance from the city center at which normalized tweet activity falls below this baseline level.

\begin{align}
R_{\text{fandom}} &= \min\{ d_i : A(M_i) \leq A_{\text{baseline}} \}
\label{eq:fandom_radius}
\end{align}
 This distance represents the spatial boundary beyond which engagement with the team is indistinguishable from background interest. Table~\ref{tab:radii} provides examples of the intermediate quantities for several teams, including the values of $A(M_1)$, $k$, $d_{\text{baseline}}$, $N$, $A_{\text{baseline}}$, and the resulting fandom radius $R_{\text{fandom}}$. These examples demonstrate how variation in baseline distance and activity levels leads to differences in the geographic reach of team fandoms.
This methodology is applied separately for each season from 2011 to 2014, as well as to the aggregated data across all four seasons, producing both season specific and overall fandom radius estimates for each NFL team.

After defining fandom regions, we analyze the sentiment of tweets associated with each fandom using the Hedonometer \cite{Dodds2011}. The Hedonometer assigns happiness scores to individual words based on human ratings of happiness. Scores range from 1 to 9, where 1 represents very negative sentiment and 9 represents very positive sentiment.

Tweets were grouped by fandom and assigned an overall happiness score based on the words they contained. To prevent certain terms from biasing sentiment, words such as team names and player names (e.g., ``Cowboys,” ``Redskins,” ``Eagles,” ``Cousins”) were removed using pattern matching with regular expressions (See Table \ref{tab:words_removed} in Appendix). After this cleaning step, tweets containing game-specific hashtags (e.g., \#NEvsBUF) were included in the datasets for both teams involved in the game. This means certain tweets were included twice, once for each team, but this double counting occurs uniformly across all games and teams, ensuring that the relative comparisons of team and fandom sentiment remain valid. Most NFL related tweets in the dataset fell within a range of approximately 5.8 to 6.2, indicating slightly positive sentiment on average, comparable to the range observed for all English tweets on \url{http://www.hedonometer.org}.

Sentiment is analyzed using two approaches. The first approach considered all tweets mentioning a given team, regardless of geographic location. For each team, tweet text containing aforementioned hashtags is aggregated and an average happiness score is calculated.
The second approach analyzed sentiment at the fandom level. In this case, only tweets originating from within the geographic radius defining a team's fandom were included. The average happiness score of these tweets is then calculated to represent the sentiment of that local fanbase.

\section{Results and Discussion}

\begin{figure*}[t]
    \centering
    \includegraphics[width=\textwidth]{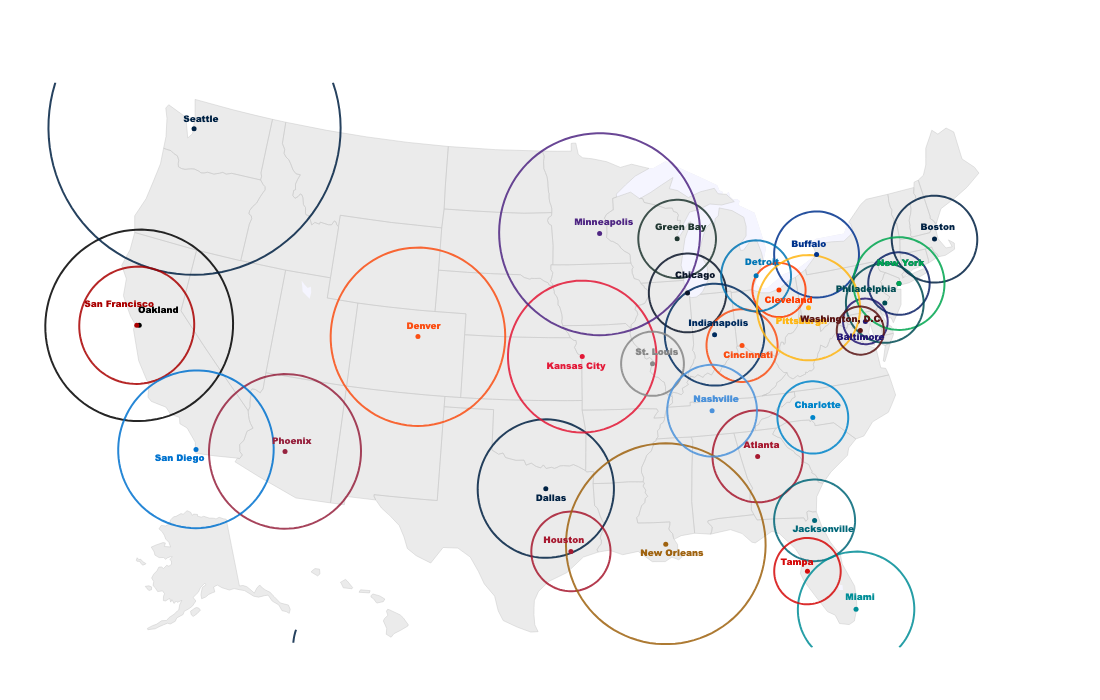}
    \caption{
        Fandom radius map of NFL teams across the United States. Each circle represents the approximate geographic reach of a team’s fanbase. Franchises located in regions with fewer nearby teams tend to have larger fandom radii, as they serve as the primary team for fans across multiple states. Conversely, teams located in densely populated regions with many nearby franchises tend to have smaller fandom radii due to competition for local fanbases.
        }
    \label{fig:us_map}
\end{figure*}

\subsection{Geographic Fandom Structure}

Calculating fandom radii provides insight into the geographic ``reach'' of each NFL team's fanbase. Larger radii indicate a more geographically dispersed fanbase, while smaller radii indicate a more concentrated fanbase centered around the team's home region.

Fandom radii vary substantially across the league, ranging from 121 kilometers for the Baltimore Ravens to 782 kilometers for the Seattle Seahawks
(see Fig.~\ref{fig:overall_engagement}).
Teams such as the Seattle Seahawks and Minnesota Vikings have the largest radii, suggesting fanbases that extend beyond their home metropolitan areas. These teams also have substantial geographic boundaries, namely the Pacific Ocean and Canada, contributing to their larger radii. In contrast, teams such as the Baltimore Ravens and Washington Redskins exhibit smaller radii, indicating more geographically concentrated fanbases. We note that the basic geographic constraints associated with national borders and oceans complicate interpretation of these rankings.

The patterns we identify appear to reflect the geographic distribution of NFL teams. Franchises located in regions with fewer nearby teams tend to have larger fandom radii, as they serve as the primary team for fans across multiple states. Conversely, teams located in densely populated regions with many nearby franchises tend to have smaller fandom radii due to competition for local fanbases.

The spatial distribution of these fandom regions is illustrated in Figure \ref{fig:us_map}. The underlying tweet distributions and fitted fandom radii for each team is provided in Appendix Figure~\ref{fig:grid_radii}. Together, these figures highlight significant overlap between several fandom regions, particularly in the Midwest and Northeast. For example, teams located in close proximity, such as Oakland and San Francisco, show overlapping fandom regions, indicating shared or competing fan engagement within the same geographic area.

\begin{figure}[t!]
    \centering    \includegraphics[width=\columnwidth]{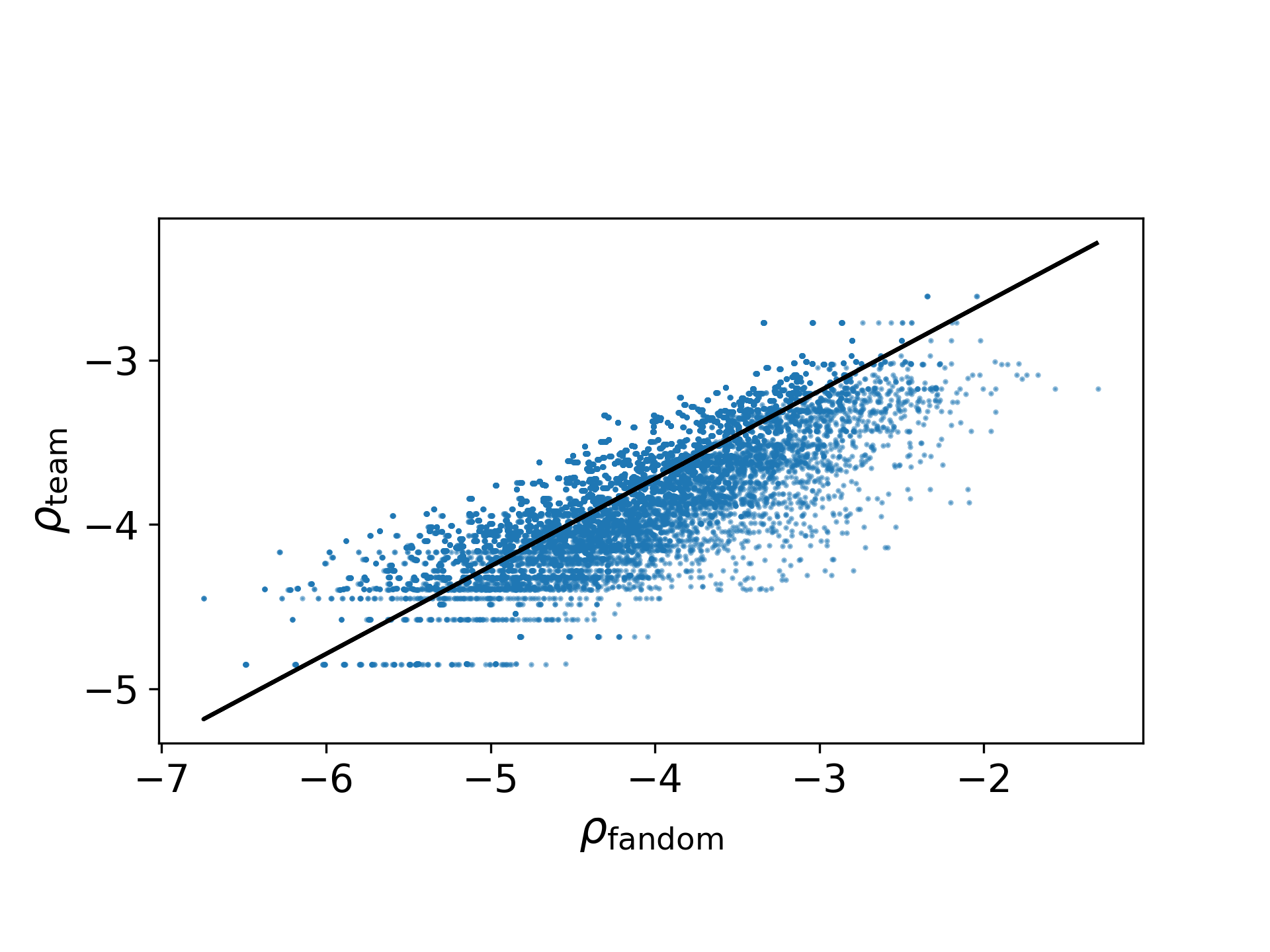}
    \caption{
    Log--log relationship between fandom density and metropolitan area size. Four seasons of data for 32 teams were represented ($\sim1500$ tweets per team per season). A total of $19,174$ data points are included, with each point representing a team--season--metropolitan area combination. The horizontal axis, $\rho_{\text{fandom}}$, is $\log_{10}(\text{tweets per 100k residents} / \text{area})$, and the vertical axis, $\rho_{\text{team}}$, is $\log_{10}(1 / \text{area})$. The black line shows the Reduced Major Axis (RMA) fit with slope $\alpha = 0.53$. The scaling relationship ($\alpha < 1$) indicates that fandom intensity increases more slowly than spatial density, consistent with a geographically dispersed fanbase.
    }
    \label{fig:density}
\end{figure}

% Density findings

Additional analysis examined how fandom density scales with metropolitan area size. Figure \ref{fig:density} shows a comparison between fandom intensity and spatial reach. The horizontal axis, $\rho_{\text{fandom}}$, is defined as $\log_{10}(\text{fandom size} / \text{area})$, where fandom size is measured as tweets per 100,000 
residents. 
The vertical axis, $\rho_{\text{team}}$, is defined as $\log_{10}(1 / \text{area})$. Each point represents a metropolitan statistical area with at least one tweet, aggregated across all teams and seasons from 2011--2014.

To quantify the relationship between these variables, we estimate $\alpha$ in
\[
\rho_{\text{team}} \propto \rho_{\text{fandom}}^{\alpha},
\]
using Reduced Major Axis (RMA) regression. The RMA slope is given by $\alpha = \mathrm{sign}(r)\,\sigma_y / \sigma_x$, where $r$ is the Pearson correlation coefficient and $\sigma_x, \sigma_y$ are the standard deviations of the respective variables.

This scaling relationship captures how fandom intensity varies with the spatial reach of metropolitan areas. We find $\alpha = 0.53$ and $R^2 = 0.64$, indicating a moderate positive correlation. An alpha value less than 1 suggests that fandom is relatively spread out, increasing more slowly than would be expected under proportional scaling with spatial density. The observed value implies that fandom is less spatially concentrated and extends beyond densely populated metropolitan areas.

\subsection{In-Game Patterns}

\begin{figure}[t!]
\centering
\includegraphics[width=\columnwidth]{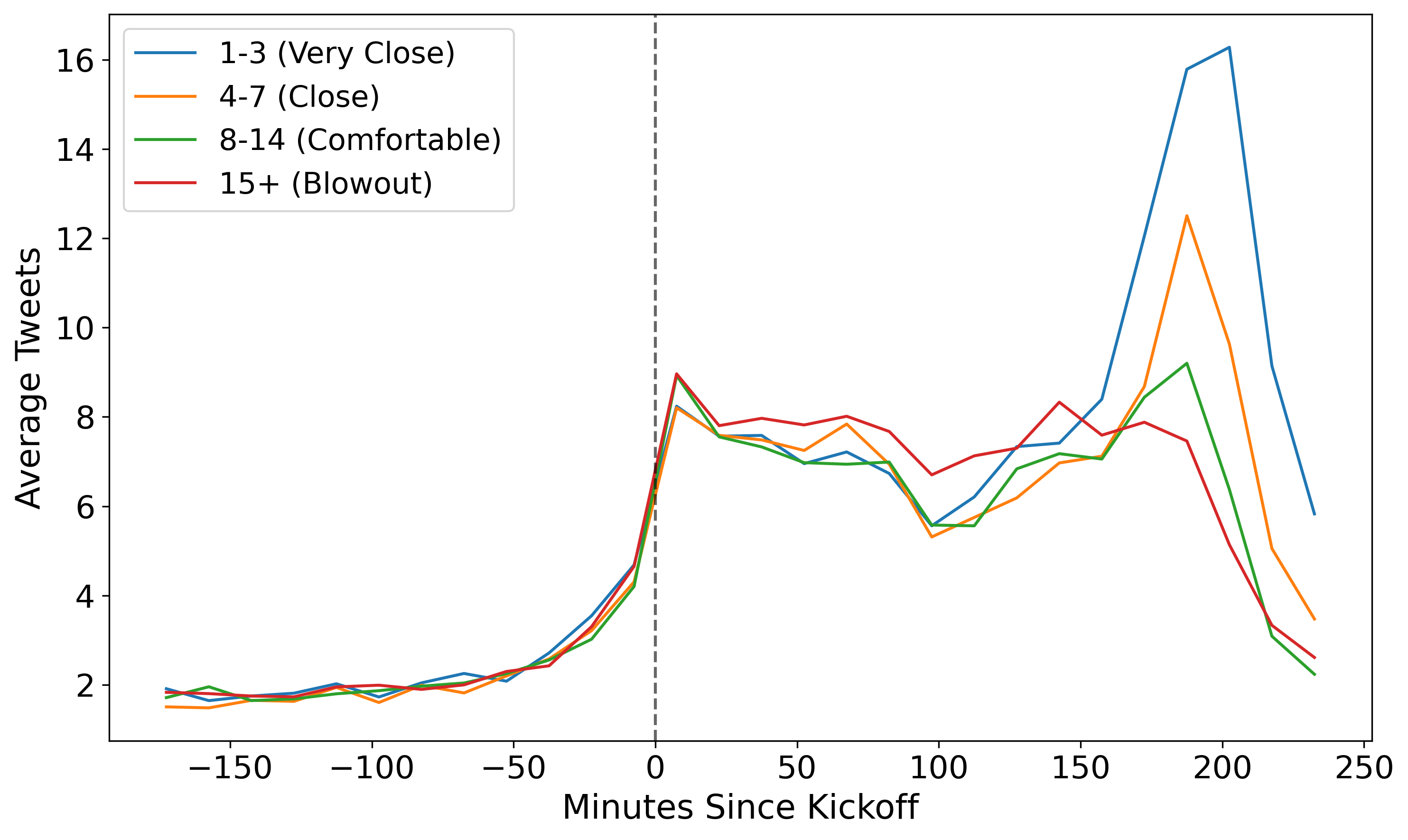}
\caption{
Average tweet volume per 15 minutes during games by point differential. Tweet counts are aggregated in 15 minute intervals from 3 hours before kickoff to 4 hours after kickoff. Lines are colored by game outcome: 1-3 points (Very Close), 4-7 points (Close), 8-14 points (Comfortable), and 15+ points (Blowout).
}
\label{fig:volume_vs_result}
\end{figure}

To understand fan engagement in real time, we analyze tweet activity for all NFL teams from three hours before kickoff to four hours after kickoff on game day.
Figure \ref{fig:volume_vs_result} shows the average tweet activity for different types of game outcomes. The goal of this analysis is to determine whether fans are more engaged during close games compared to games where the outcome is relatively certain. Across all teams, three key moments stand out: kickoff, halftime, and game end. Tweet activity generally spikes at kickoff, dips during halftime (typically around 100 minutes after kickoff), and reaches a larger peak at the conclusion of the game (roughly three hours after kickoff \cite{Kennedy2025}).

\begin{figure*}[t!]
    \centering
    \includegraphics[width=\textwidth]{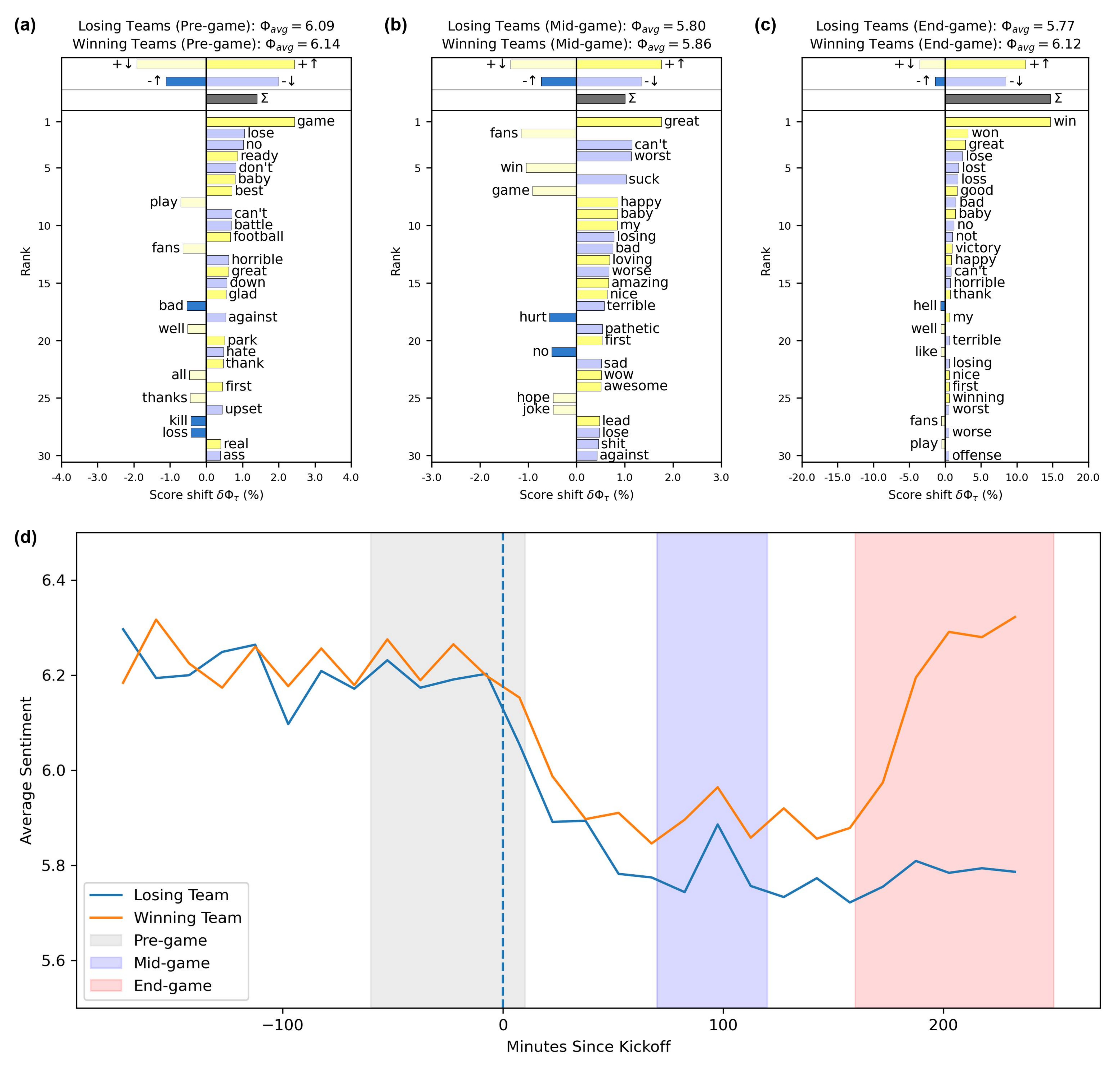}
    \caption{Comparison of word shifts and tweet sentiment for the eventual winning and losing teams during the game. Panels (a)-(c) show word shifts at three time windows relative to kickoff: (a) 60 minutes before the game until 10 minutes after kickoff, (b) 70 minutes after kickoff until 120 minutes after kickoff, and (c) 160 minutes after kickoff until 250 minutes after kickoff. Dark yellow bars represent positive words used more frequently by the winning team and less frequently by the losing team, light blue bars represent negative words used less frequently by the winning team and more frequently by the losing team, light yellow bars represent positive words used less frequently by the winning team, and dark blue bars represent negative words used more frequently by the winning team. Panel (d) shows the in-game sentiment trajectory for winning and losing teams (aggregated in 15-minute intervals) relative to kickoff, with colored sections matching the corresponding word shift panels (a)--(c).}
    \label{fig:in_game_sent}
\end{figure*}

\begin{table*}[t!]
\centering
\setlength{\tabcolsep}{3pt}
\scriptsize

\begin{ruledtabular}
\begin{tabular}{lrrrrrr}
Team & Radius (km) & Total Mentions & Mentions (Inside) & \% Inside Radius & Happiness (Inside) & Happiness (Total) \\
\hline
Arizona Cardinals & 411.50 & 2078 & 203 & 9.77 & 5.95 & 5.97 \\
Atlanta Falcons & 244.78 & 4047 & 1338 & 33.06 & 5.95 & 5.95 \\
Baltimore Ravens & 121.17 & 4948 & 1653 & 33.41 & 5.97 & 5.93 \\
Buffalo Bills & 229.57 & 3972 & 1586 & 39.93 & 5.96 & 5.94 \\
Carolina Panthers & 192.10 & 3445 & 1095 & 31.79 & 6.09 & 6.02 \\
Chicago Bears & 209.78 & 8052 & 3184 & 39.54 & 5.81 & 5.86 \\
Cincinnati Bengals & 193.15 & 2884 & 1016 & 35.23 & 5.99 & 5.98 \\
Cleveland Browns & 144.26 & 4691 & 1735 & 36.99 & 5.95 & 5.90 \\
Dallas Cowboys & 368.69 & 12565 & 2876 & 22.89 & 6.03 & 5.97 \\
Denver Broncos & 474.04 & 7088 & 1391 & 19.62 & 6.08 & 6.04 \\
Detroit Lions & 189.24 & 5207 & 1454 & 27.92 & 5.90 & 5.95 \\
Green Bay Packers & 209.61 & 8028 & 1633 & 20.34 & 5.99 & 5.96 \\
Houston Texans & 213.10 & 4617 & 2241 & 48.54 & 5.94 & 5.96 \\
Indianapolis Colts & 270.58 & 3854 & 1292 & 33.52 & 6.10 & 6.04 \\
Jacksonville Jaguars & 218.16 & 1495 & 529 & 35.38 & 6.06 & 5.98 \\
Kansas City Chiefs & 403.40 & 4476 & 1978 & 44.19 & 5.97 & 5.94 \\
Miami Dolphins & 310.20 & 3352 & 851 & 25.39 & 6.10 & 6.01 \\
Minnesota Vikings & 541.69 & 3720 & 1395 & 37.50 & 5.94 & 5.98 \\
New England Patriots & 231.55 & 7800 & 2223 & 28.50 & 5.98 & 5.96 \\
New Orleans Saints & 537.33 & 5027 & 1333 & 26.52 & 6.11 & 6.03 \\
New York Giants & 166.41 & 6347 & 1993 & 31.40 & 5.97 & 5.99 \\
New York Jets & 246.87 & 5418 & 2412 & 44.52 & 5.88 & 5.88 \\
Oakland Raiders & 509.26 & 2984 & 789 & 26.44 & 5.99 & 5.96 \\
Philadelphia Eagles & 210.93 & 8359 & 3741 & 44.75 & 5.93 & 5.95 \\
Pittsburgh Steelers & 279.76 & 7947 & 3070 & 38.63 & 5.97 & 5.96 \\
Seattle Seahawks & 781.69 & 6160 & 2509 & 40.73 & 6.08 & 6.05 \\
San Francisco 49ers & 311.85 & 8015 & 2696 & 33.64 & 6.02 & 6.00 \\
San Diego Chargers & 420.35 & 3208 & 1191 & 37.13 & 6.12 & 6.03 \\
St. Louis Rams & 170.58 & 2067 & 343 & 16.59 & 5.90 & 5.95 \\
Tampa Bay Buccaneers & 178.15 & 1127 & 229 & 20.32 & 5.83 & 5.93 \\
Tennessee Titans & 243.61 & 1995 & 639 & 32.03 & 5.85 & 5.90 \\
Washington Redskins & 127.89 & 4361 & 1426 & 32.70 & 5.90 & 5.89 \\
\end{tabular}
\end{ruledtabular}

\caption{Geographic fandom radius and sentiment comparison for NFL teams. Each row reports the estimated fandom radius (km), total tweet volume, the number and percentage of tweets originating within that radius, and average Hedonometer sentiment scores inside the radius versus overall. Since our data is a 10\% random sample, raw counts in the Mentions columns should be multiplied by 10 to approximate the actual number of messages.}
\label{tab:fandom_radius_sentiment}

\end{table*}

For this analysis, games were grouped into four buckets: very close games, where the score difference is between 1 and 3 points (within a field goal of tying or winning), close games, with a 4 to 7 point margin (within a touchdown), comfortable wins with margins between 8 and 14 points, and blowouts, where a team led by at least 15 points.

Close games tend to spike around 150 minutes after kickoff (around the fourth quarter), which is consistent with the idea that fans become more attentive when the outcome is uncertain. Interestingly, blowout games show slightly higher average tweet volume during portions of the game. One possible explanation is that these games may include multiple scoring events or highlight worthy plays that prompt online discussion even when the overall outcome is less competitive.
Note that the point differential used in this analysis reflects the final score margin, rather than the margin at each moment in the game. As a result, some games classified as blowouts may have been competitive earlier before one team pulled away late.

Moving to emotional reactions, Figure \ref{fig:in_game_sent} provides insight into the sentiment differences between tweets about winning and losing teams throughout a game. The top panel shows word shift graphs for both the winning and losing teams at three different stages of a game \cite{Gallagher2021}. The bottom panel displays fan sentiment throughout a game for the eventual winning and losing teams.

Sentiment for tweets referencing winning and losing teams is very similar before kickoff, and both measures decline noticeably once the game begins. This pattern likely reflects pre-game optimism from both fanbases, followed by more critical or emotional reactions once the game begins. Throughout the game, sentiment for the eventual winning team remains slightly higher than that of the losing team.

Around the 100 minute mark, corresponding roughly to halftime, sentiment increases slightly for both groups. This increase may occur because fans temporarily shift discussion toward other NFL related topics or general game commentary during the break in game play. After halftime, sentiment begins to diverge more clearly. Sentiment associated with the winning team gradually increases, while sentiment for the losing team remains relatively stable. This pattern suggests that fans of the winning team increasingly express positive reactions as the outcome becomes more certain, whereas fans of the losing team continue to express more neutral or negative emotions.

Pre-game tweets show similar sentiment, with the winning team at 6.14 and the losing team at 6.09, reflecting general excitement; common positive words include ``game'', ``ready'', ``play'', and ``fans.'' During halftime, sentiment drops slightly to 5.86 for the winning team and 5.80 for the losing team. By the end of the game, the difference is more pronounced, with the winning team at 6.12 and the losing team at 5.77. These differences are driven by the greater use of positive words such as ``win'', ``great'', and ``victory'' in tweets about winning teams, as well as the more frequent use of negative words such as ``lose'', ``bad'', and ``horrible'' for losing teams. The four color summary bars at the top of the graph indicate that the dominant contributors to the overall difference are an increase in positive language, followed by a decrease in negative language in winning team tweets. Words going against this trend, namely relatively happy words used less often in winning team tweets and relatively sad words used more often in winning team tweets, are far less common.

These results support a hypothesis popularized by Kurt Vonnegut that the stories we love to tell exhibit a small handful of emotional arcs \cite{Vonnegut2010}. Fans of both teams begin the game with optimism, posting tweets whose average happiness is on par with the best individual days for all of Twitter as measured by the Hedonometer. As the game begins, fans of both winning and losing teams experience plummeting sentiment. Consistent with Vonnegut's ``Man in a Hole'' emotional arc, fans of winning teams subsequently experience an emotional rise beginning near the start of the fourth quarter and ending in triumph at a sentiment slightly higher than the optimistic pre-game excitement \cite{Vonnegut2010}. The losing team on the other hand experiences Vonnegut's ``Tragedy'' emotional arc, exemplified by Romeo and Juliet, and finishing with a sentiment comparable to the saddest days for all of Twitter as measured by the Hedonometer. Future work might explore the relationship between the shapes of individual game scores and fan experience \cite{Kiley2016}.

\subsection{Sentiment Variations}

In addition to comparing the sentiment of tweets from winning and losing teams, sentiment is analyzed for tweets posted within a team's fandom radius compared to all tweets posted. Table~\ref{tab:fandom_radius_sentiment} summarizes tweet activity and sentiment for each NFL team. Specifically, the table includes the total number of tweets associated with each team, the number and percentage originating within the fandom radius, and the average sentiment both within the radius and across all tweets. Comparing these values helps reveal whether fans located near a team’s geographic fanbase express sentiment differently than the broader online audience.

Table~\ref{tab:fandom_radius_sentiment} shows considerable variation across the league in the percentage of tweets posted within a team’s fandom radius. For some teams, such as the Houston Texans, Kansas City Chiefs, and Philadelphia Eagles, over 40\% of tweets originate within this radius, suggesting a strong geographic concentration of fan activity. In contrast, teams such as the Arizona Cardinals and St. Louis Rams have fewer than 20\% of tweets within their radius, indicating that discussion of these teams is more geographically dispersed.

This variation is expected given how the fandom radius is defined. The radius represents the geographic boundary within which fan engagement exceeds background levels, rather than the full extent of the fanbase, and is intentionally conservative. It is calculated as the distance at which normalized tweet activity falls to baseline, capturing only the densest region of above background engagement. Consequently, a low percentage of tweets within the radius does not indicate low overall fan activity, but rather reflects the broader spatial distribution of a team’s fans. In other words, teams with a low percentage of tweets inside the radius may have many distant fans whose activity contributes to baseline levels, rather than extending the radius itself.

\begin{figure}[t!]
    \centering
    \includegraphics[width=0.85\columnwidth]{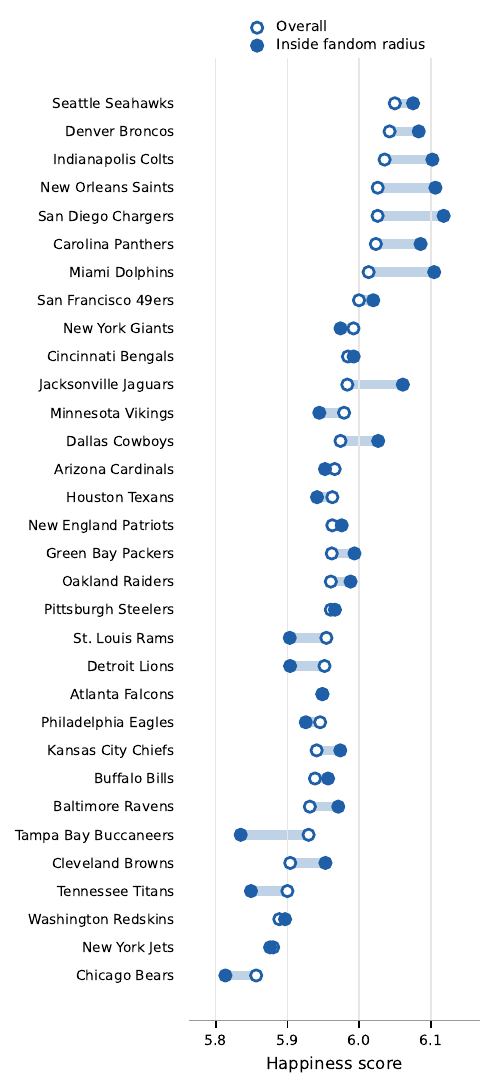}
    \caption{Comparison of happiness scores for each NFL team. Each row connects two values with a bar: the average happiness score of tweets posted inside the team's fandom radius (filled circle) and the average happiness score of all tweets mentioning the team, regardless of location (open circle). Teams are sorted by overall happiness score.}
    \label{fig:sentiment_comparison}
\end{figure}

\begin{figure}[t]
    \centering
    \includegraphics[width=\columnwidth]{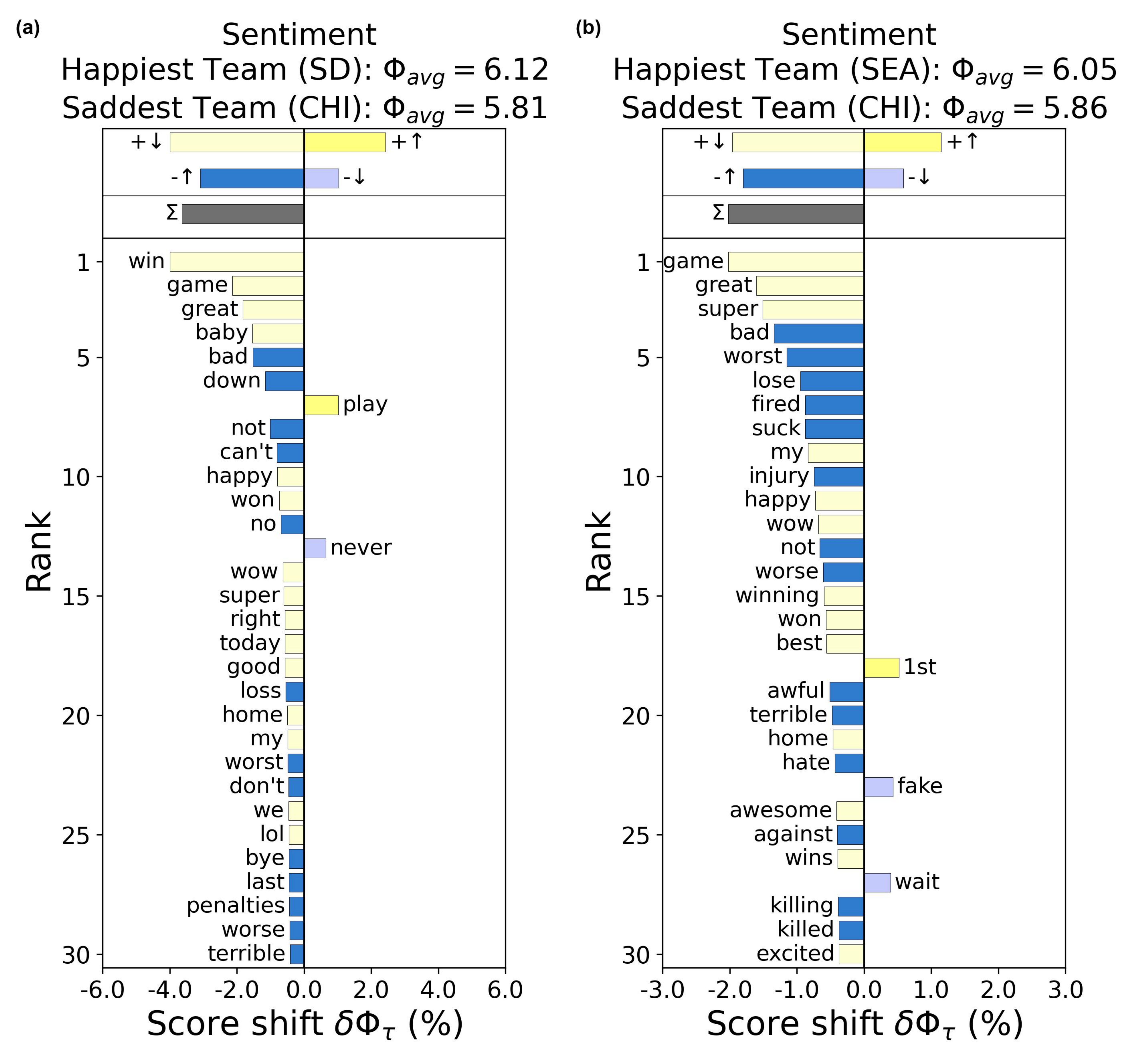}
    \caption{Comparison of word shifts for the happiest and saddest teams, inside the fandom radius and across all tweets~\cite{Gallagher2021}. (a) Tweets written inside the fandom radius, comparing the happiest team (San Diego Chargers) and saddest team (Chicago Bears). (b) All tweets posted, comparing the happiest team (Seattle Seahawks) and saddest team (Chicago Bears). Light yellow bars represent positive words used more frequently by the happiest team and less frequently by the saddest team. Dark blue bars represent negative words used more frequently by the saddest team and less frequently by the happiest team. Dark yellow bars represent positive words used more frequently by the saddest team, while light blue bars represent negative words used less frequently by the saddest team and more frequently by the happiest team.}
    \label{fig:wordshift_comparison} 
\end{figure}

Despite these differences in geographic concentration, the average sentiment of tweets posted inside a fandom region compared to the overall sentiment remains relatively similar. This suggests that location is not a major factor influencing sentiment variation in tweets. Fans tend to express similar tones when composing tweets regardless of where they are located. The differences between sentiment inside the fandom radius and overall sentiment are generally small across teams, indicating that the emotional tone of online discussions is relatively consistent across geographic regions.

Figure~\ref{fig:sentiment_comparison} compares the average sentiment of tweets for each NFL team within their fandom radius (filled circles) with overall sentiment across all tweets (open circles). Sentiment scores for tweets inside fandom regions range from approximately 5.81 to 6.12, while overall sentiment across all tweets spans a slightly narrower range of roughly 5.86 to 6.05. In both groupings, the Chicago Bears consistently exhibit the lowest sentiment, whereas the highest sentiment is observed for the San Diego Chargers within fandom regions and the Seattle Seahawks across all tweets. The Spearman correlation between fandom region sentiment and overall sentiment is $\rho = 0.85$ $(p < 0.001)$, indicating a strong positive relationship. The close similarity in sentiment patterns between the two plots indicates that, for most teams, the geographic location of the fan has little influence on the overall emotional tone of tweets.

Figure~\ref{fig:wordshift_comparison} shows two word shift graphs comparing the language used by the happiest and saddest teams, both within their fandom radius and across all tweets~\cite{Gallagher2021}. The graph on the left compares tweets written inside the fandom radius for the San Diego Chargers and the Chicago Bears. Positive words such as “win,” “game,” and “great” appear less frequently in Bears tweets, while negative words such as “bad” and “down” appear more frequently, contributing to the lower sentiment score for Bears tweets. The interpretation of some words is less straightforward; for example, the word “down” may refer to a football down rather than negative sentiment depending on context. Interestingly, the word “play” appears more frequently in Bears tweets and less frequently in Chargers tweets, while the word “never” appears less frequently in Bears tweets and more frequently in Chargers tweets.

The graph on the right compares overall tweets for the Seattle Seahawks and the Chicago Bears. Bears tweets contain positive words such as “game,” “great,” and “super” less frequently than Seahawks tweets, which contributes to the lower overall sentiment score. Bears tweets also include more negative words such as “bad,” “worst,” “lose,” “fired,” and “suck,” further lowering their sentiment relative to Seahawks tweets. Overall, these word shifts provide insight into the differences in sentiment between fandoms by highlighting the relative frequency of positive and negative game related language.

\begin{figure}[t!]
    \centering
    \includegraphics[width=\columnwidth]{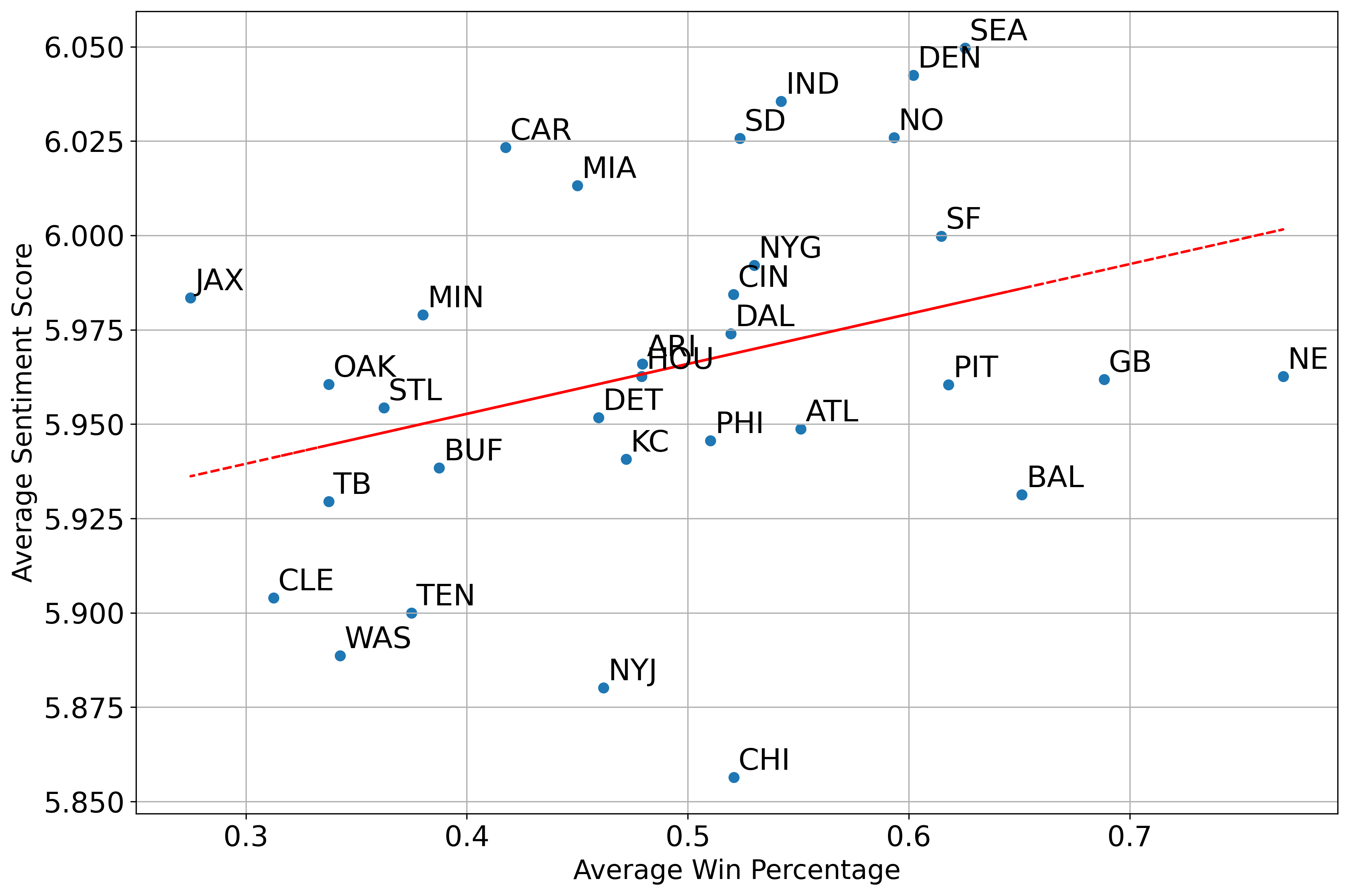}
    \caption{Relationship between average team win percentage (2011–2014) and average tweet sentiment. Each point represents an NFL team, with a regression line indicating the overall trend.}
    \label{fig:sent_winpct}
\end{figure}

Another component of this analysis examined whether a team’s record, measured by winning percentage, offered predictive utility in estimating tweet sentiment. Figure~\ref{fig:sent_winpct} compares each team’s average sentiment score from 2011 to 2014 with their average win percentage over the same period. The scatter plot shows a weak positive correlation between winning percentage and sentiment, indicating that teams with higher win percentages tend to have slightly higher sentiment scores. A linear regression line illustrates this upward trend, and the Pearson correlation between win percentage and sentiment score is $0.33$. This finding suggests that while winning is associated with positive tweet sentiment, team performance alone does not fully explain observed sentiment differences across teams. Other factors, such as game context, geographic variation in fanbases, team culture, and fan expectations, may also influence how fans express sentiment on Twitter.

\section{Discussion}

Using Twitter data to estimate fandom locations and analyze fan sentiment provides a new perspective on social media engagement among NFL fans. Our analysis reveals substantial variation in fandom radii across the league. Teams located in regions without nearby NFL teams tend to have larger fandom radii, while teams located in dense clusters of franchises often have smaller, overlapping radii. This research also provides insight into the sentiment of different fandoms and allowed for the quantification of relatively ``happy” and ``sad” fanbases. Sentiment varied across teams, and while the range of scores differed slightly depending on whether sentiment is calculated overall or only within the fandom radius, consistent groupings of happier, neutral, and less happy teams emerged.

Examining tweeting patterns during games also provided insight into how fans respond to game events. Tweet activity tended to spike shortly after kickoff and again toward the end of games, reflecting moments of heightened fan attention. Winning teams also generated slightly more tweets than losing teams, suggesting that fans of losing teams may be less inclined to share reactions online. Sentiment differences between winning and losing teams were especially visible near the end of games, which aligns with expectations that fans of winning teams are more likely to express positive reactions, often using celebratory language.

Social media platforms allow fans to react to games in real time regardless of their geographic location. The results of this research support the idea that fan emotion on Twitter is driven more strongly by in game events and final outcomes than by geographic location. Social media sentiment also has the potential to help measure fan engagement in real time. Geographic engagement patterns can help teams identify the difference between local and more widely distributed fanbases. Sentiment analysis can also provide insight into fan reactions to games, players, coaching decisions, and officiating decisions. These data could complement traditional measures of fan engagement such as television ratings and stadium attendance. 

This study has several limitations. The analysis is limited to the seasons with the most available Twitter data (2011–2014) and only includes a 10\% sample of tweets. Additionally, the analysis relied only on geolocated tweets, which represent a small subset of Twitter users. Many fans may not use Twitter to discuss sports or may not engage in real-time conversations during games due to factors such as age, internet access, or cultural preferences. The Hedonometer also evaluates words individually, which can miss contextual meaning such as sarcasm or football specific terminology. Furthermore, tweets were collected using specific game related hashtags, which excludes tweets that discuss games without hashtags, use alternative hashtags, or contain misspellings.

Future research could expand this work by examining how fans react to different types of games. One potential direction is to compare pre-game betting odds and point spreads with tweet sentiment to determine whether fan sentiment reflects expectations about which team is likely to win. Additional game level analysis could also examine how sentiment changes in response to specific scoring events during games to better understand which moments generate the strongest emotional reactions among fans. Social media data could further be used to identify which game features, such as close scores, impressive plays, or star player performance, drive the highest levels of attention and engagement.

\bigskip
\bigskip

\acknowledgments

The authors are grateful for support from the
National Science Foundation (Award \#2242829),
the Massachusetts Mutual Life Insurance Company,
and an anonymous philanthropic donor.

\bibliographystyle{apsrev4-2}
\bibliography{references}

\clearpage
\FloatBarrier 
\appendix

\onecolumngrid

\appendix

\renewcommand{\thesection}{A}
\renewcommand{\thepage}{A\arabic{page}}
\renewcommand{\thefigure}{A\arabic{figure}}
\renewcommand{\thetable}{A\arabic{table}}
\setcounter{section}{0}
\setcounter{page}{1}
\setcounter{figure}{0}
\setcounter{table}{0}

\section{Additional Tables and Figures} 

\begin{table}[!htbp]
\centering
\scriptsize

\begin{tabular}{|l|l|l|}
\hline
\textbf{Team} & \textbf{Home City} & \textbf{Hashtags} \\
\hline
Arizona Cardinals & Phoenix, AZ & \#ArizonaCardinals, \#Cardinals \\
\hline
Atlanta Falcons & Atlanta, GA & \#AtlantaFalcons, \#Falcons \\
\hline
Baltimore Ravens & Baltimore, MD & \#BaltimoreRavens, \#Ravens \\
\hline
Buffalo Bills & Buffalo, NY & \#BuffaloBills, \#Bills \\
\hline
Carolina Panthers & Charlotte, NC & \#CarolinaPanthers, \#Panthers \\
\hline
Chicago Bears & Chicago, IL & \#ChicagoBears, \#Bears \\
\hline
Cincinnati Bengals & Cincinnati, OH & \#CincinnatiBengals, \#Bengals \\
\hline
Cleveland Browns & Cleveland, OH & \#ClevelandBrowns, \#Browns \\
\hline
Dallas Cowboys & Dallas, TX & \#DallasCowboys, \#Cowboys \\
\hline
Denver Broncos & Denver, CO & \#DenverBroncos, \#Broncos \\
\hline
Detroit Lions & Detroit, MI & \#DetroitLions, \#Lions \\
\hline
Green Bay Packers & Green Bay, WI & \#GreenBayPackers, \#Packers \\
\hline
Houston Texans & Houston, TX & \#HoustonTexans, \#Texans \\
\hline
Indianapolis Colts & Indianapolis, IN & \#IndianapolisColts, \#Colts \\
\hline
Jacksonville Jaguars & Jacksonville, FL & \#JacksonvilleJaguars, \#Jaguars \\
\hline
Kansas City Chiefs & Kansas City, MO & \#KansasCityChiefs, \#Chiefs \\
\hline
Miami Dolphins & Miami, FL & \#MiamiDolphins, \#Dolphins \\
\hline
Minnesota Vikings & Minneapolis, MN & \#MinnesotaVikings, \#Vikings \\
\hline
New England Patriots & Boston, MA & \#NewEnglandPatriots, \#Patriots \\
\hline
New Orleans Saints & New Orleans, LA & \#NewOrleansSaints, \#Saints \\
\hline
New York Giants & New York, NY & \#NewYorkGiants, \#Giants \\
\hline
New York Jets & New York, NY & \#NewYorkJets, \#Jets \\
\hline
Oakland Raiders & Oakland, CA & \#OaklandRaiders, \#Raiders \\
\hline
Philadelphia Eagles & Philadelphia, PA & \#PhiladelphiaEagles, \#Eagles \\
\hline
Pittsburgh Steelers & Pittsburgh, PA & \#PittsburghSteelers, \#Steelers \\
\hline
Seattle Seahawks & Seattle, WA & \#SeattleSeahawks, \#Seahawks \\
\hline
San Francisco 49ers & San Francisco, CA & \#SanFrancisco49ers, \#Niners \\
\hline
San Diego Chargers & San Diego, CA & \#SanDiegoChargers, \#Chargers \\
\hline
St. Louis Rams & St. Louis, MO & \#StLouisRams, \#Rams \\
\hline
Tampa Bay Buccaneers & Tampa, FL & \#TampaBayBuccaneers, \#Buccaneers \\
\hline
Tennessee Titans & Nashville, TN & \#TennesseeTitans, \#Titans \\
\hline
Washington Redskins & Washington, D.C. & \#WashingtonRedskins, \#Redskins \\
\hline
\end{tabular}

\caption{NFL teams, their home cities, and associated hashtags used for tweet querying.}
\label{tab:hashtags}

\end{table}

\begin{table}[b!]
\centering
\scriptsize

\begin{tabular}{|l|}
\hline
\textbf{Removed Words} \\
\hline
cowboys \\
\hline
redskins \\
\hline
cousins \\
\hline
washington \\
\hline
eagles \\
\hline
lynch \\
\hline
saints \\
\hline
bills \\
\hline
love \\
\hline
birds \\
\hline
fly \\
\hline
eagle \\
\hline
sea \\
\hline
miami \\
\hline
sun \\
\hline
beach \\
\hline
rivers \\
\hline
luck \\
\hline
golden \\
\hline
high \\
\hline
beast \\
\hline
\end{tabular}

\caption{Words removed from tweets to prevent bias in sentiment analysis.}
\label{tab:words_removed}

\end{table}

\FloatBarrier

%% \section{Additional Figures}

\begin{figure} [t]
    \centering
    \includegraphics[width=\textwidth]{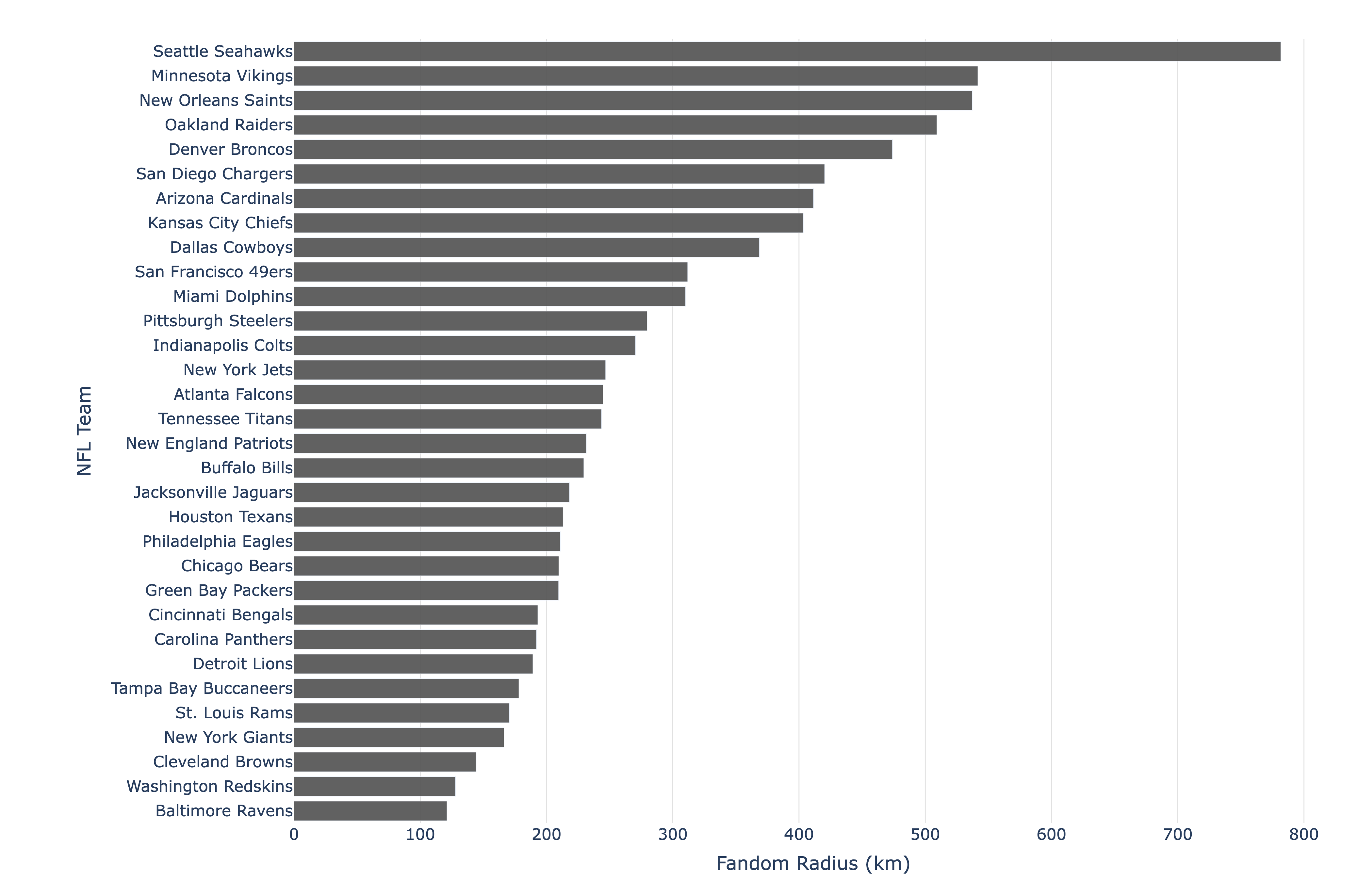}
    \caption{
        Estimated fandom radius for each NFL team calculated by using geotagged Twitter activity. The radius reflects the geographic distance required to capture the majority of fan tweets associated with a team, with larger values indicating more geographically dispersed fanbases.
    }
    \label{fig:overall_engagement}
\end{figure}

\begin{figure}
    \centering
    \includegraphics[width=\linewidth]{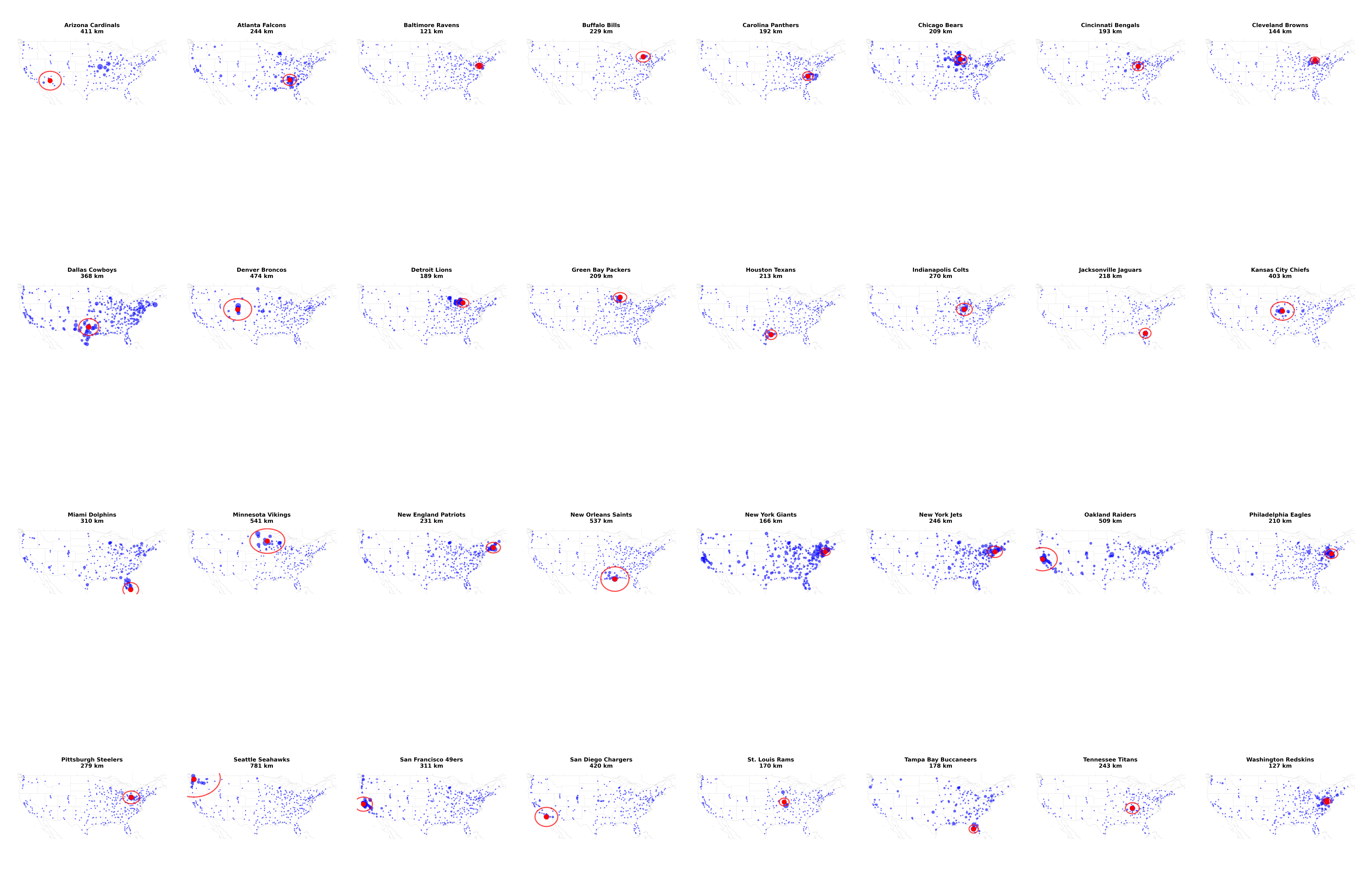}
    \caption{
        Estimated fandom radii and geographic tweet activity for all 32 NFL teams. Each team’s home city is marked with a red dot and its fandom radius is shown as a red circle. Blue dots indicate tweet activity in metro areas, with larger dots corresponding to higher tweet volume per 100,000 residents.
    }
    \label{fig:grid_radii}
\end{figure}

\twocolumngrid

\end{document}